\documentclass[
aps,
superscriptaddress,nofootinbib,
reprint,
]{revtex4-2}

\usepackage{xcolor}
\usepackage{graphicx}
\usepackage{amsmath,amssymb,amsfonts,mathrsfs,bm}
\usepackage[
colorlinks=true,
linkcolor=blue,
citecolor=blue,
]{hyperref}

\def\l{\left}
\def\r{\right}
\def\ddd{\mathrm{d}}

\begin{document}
	
\title{Bridging inflation and reheating: chiral gravitational waves from aHz to GHz}

\author{ Chengjie Fu}
\email[]{fucj@ahnu.edu.cn}
\affiliation{ Department of Physics, Anhui Normal University, Wuhu, Anhui 241002, China }

\author{ Chao Chen}
\email[Corresponding author:~]{cchao012@just.edu.cn}
\affiliation{School of Science, Jiangsu University of Science and Technology, Zhenjiang, 212100, China }

\author{ Yi Wang } 
\email[]{phyw@ust.hk}
\affiliation{Department of Physics, The Hong Kong University of Science and Technology,
Clear Water Bay, Kowloon, Hong Kong, China}
\affiliation{ Jockey Club Institute for Advanced Study, The Hong Kong University of Science and Technology, Clear Water Bay, Kowloon, Hong Kong, China}

\begin{abstract}

In this paper, we investigate chiral gravitational wave (GW) signals generated from inflation to reheating, driven by a parity-violating (PV) term coupled to the inflaton,
$\dot\phi\epsilon^{ijk} h_{il}\partial_j h_{k}^l$, which naturally arises in PV extensions of teleparallel gravity.
During inflation, the PV term reduces the sound horizon for right-handed circularly polarized GWs, and amplifies their power spectra relative to left-handed GWs. At CMB scales, these chiral GWs induce BB as well as non-vanishing EB and TB correlations in CMB, which are potentially detectable by LiteBIRD. During reheating, subhorizon modes undergo tachyonic instability, leading to fully circularly polarized GWs with enhanced amplitudes, which may be probed by future high-frequency GW experiments, such as resonant cavity. The absence of backreaction effect of enhanced chiral GWs imposes constraints on the energy scale of the PV term, the inflationary potential, and the reheating history. Our findings highlight the potential of multi-frequency GW experiments to offer a unique probe of the parity violation and early Universe.

\end{abstract}

\maketitle

\section{Introduction}
The observation of parity violation in the weak interaction has already demonstrated that Nature does distinguish between the right- and left-handed chiralities of matter fields~\cite{Lee:1956qn,Wu:1957my}. 
While the most successful gravity theory---Einstein's General Relativity (GR) conserves parity, exploring the potential parity violation in gravity could significantly deepen our understanding of the nature of gravity and the Universe.
Remarkably, the recent observations of cosmic birefringence in cosmic microwave background radiation (CMB)~\cite{Minami:2020odp,Diego-Palazuelos:2022dsq} and the large-scale structure in the Universe~\cite{Hou:2022wfj,Philcox:2022hkh}\footnote{In contrast, Ref.~\cite{Krolewski:2024paz} reported no evidence of the parity violation in the large-scale structure, based on their analysis of BOSS data.} have revealed potential evidences of the parity violation related to gravity, sparking significant interest in investigating parity violation, e.g., Refs.~\cite{Cabass:2022rhr,Cai:2022lec,Zhu:2023lhv,Coulton:2023oug,Garcia-Saenz:2023zue,Jazayeri:2023kji,Stefanyszyn:2023qov,Stefanyszyn:2024msm,Inomata:2024ald,Jamieson:2024mau,Akama:2024bav,Hassan:2024vjg,Zhang:2024vfw,Jiang:2024woi}.

A wide array of parity-violating (PV) theories of gravity have been proposed in the literature, including Chern-Simons modified gravity~\cite{Jackiw:2003pm,Alexander:2009tp}, ghost-free PV scalar-tensor theory~\cite{Crisostomi:2017ugk,Nishizawa:2018srh,Gao:2019liu}, PV extensions of teleparallel gravity~\cite{Li:2020xjt,Li:2022mti} and symmetric teleparallel gravity~\cite{Conroy:2019ibo,Li:2021mdp,Li:2022vtn}, among others. 
If a parity symmetry of gravity is violated, the left- and right-handed polarization modes of gravitational waves (GWs) will follow distinct equation of motion (EoM), a phenomenon known as birefringence.
Numerous tests on birefringence of GWs have been conducted using GW data released by the LIGO-Virgo collaboration~\cite{Nishizawa:2018srh,Conroy:2019ibo,Wang:2020cub,Yamada:2020zvt,Okounkova:2021xjv,Wang:2021gqm,Wu:2021ndf,Gong:2021jgg,Ng:2023jjt}. These studies have not yet identified any conclusive signatures of parity violation. 
Since the energy scale of inflation far exceeds that accessible by current experiments, exploring PV gravity theories during inflation provides a unique opportunity to probe fundamental physics at extreme energy scales.

During inflation, PV gravitational interactions can generate circular-polarized primordial GWs~\cite{Lue:1998mq,Alexander:2004us,Satoh:2007gn,Alexander:2016hxk,Bartolo:2017szm,Qiao:2019hkz,Fu:2020tlw,Cai:2021uup,Odintsov:2022hxu,Peng:2022ttg,Fu:2023aab}. Chiral GW signals imprint distinctive features  on the temperature and polarization anisotropies of the CMB, notably inducing non-vanishing cross-correlation spectra between the CMB temperature and B-mode (TB), as well as between E-mode and B-mode (EB), alongside the B-mode auto-correlation spectrum (BB). Precise measurements of the TB and EB spectra from future CMB experiments could offer compelling evidence for parity violation in gravity~\cite{Saito:2007kt,Gluscevic:2010vv}. Furthermore, the chiral GW signals can arise from the excited gauge field in the axion inflation, where the axion couples to the gauge field through the Chern-Simons term~\cite{Barnaby:2011qe,Garcia-Bellido:2016dkw,Ozsoy:2020ccy,Almeida:2020kaq,Ozsoy:2020kat}.

In this paper, we explore the chiral primordial GWs arising from PV gravitational interactions coupled to an inflaton, $\dot\phi\epsilon^{ijk} h_{il}\partial_j h_{k}^l$, which naturally arises in PV extensions of teleparallel gravity~\cite{Li:2020xjt,Li:2022mti}. For the first time, we analyze the full evolution of chiral GWs from inflation through reheating, showing that the resulting chiral GWs get enhanced and span a broad frequency range, extending from aHz to GHz. This enables us to explore the PV coupling, the dynamics of the inflaton, and the reheating history by combining the observations of the CMB's BB, EB, and TB spectra with high-frequency GW experiments. However, current experimental sensitivities in the MHz–GHz range remain above the bound from Big Bang Nucleosynthesis \cite{Aggarwal:2025noe}, and thus these signals should be regarded as targets for future detection concepts. Our findings indicate that multi-band GW detector data will serve as a powerful tool for probing the early Universe and testing new physics.


\section{Chiral primordial GWs from inflation to reheating}
In this paper, we focus on the linear chiral gravitational waves which can generate from four possible gravitational parity-odd terms, namely $\epsilon^{ijk} \dot{h}_{i}^{l} \partial_j\dot{h}_{kl}$, $\epsilon^{ijk} \partial^2h_{il} \partial_jh_{k}^{l}$, $\epsilon^{ijk} \dot{h}_{il} \partial_jh_{k}^{l}$, $\epsilon^{ijk} h_{il}\partial_j h_{k}^l$, where $\epsilon^{ijk}$ is the totally-asymmetric symbol. See Supplementary Material for details. 
Among these, the last operator has the lowest mass dimension and is therefore the least suppressed in any low-energy effective theory. It is also free from the kinetic and gradient instability.
Hence, the quadratic action for chiral GWs is written as,   
\begin{align} \label{eq:action}
    S_{T}^{(2)} = \frac{M_{\mathrm{Pl}}^2}{8} \int\ddd t~\ddd^3x~a^3(t)&\Big[ \dot{h}_{ij}\dot{h}^{ij}-\frac{1}{a^2} \l( \partial h_{ij} \r)^2 \nonumber
    \\& \quad
    - \frac{\dot{\phi}}{a \Lambda} \epsilon^{ijk} h_{il}\partial_j h_{k}^l \Big] ~,
\end{align}
where $M_{\rm Pl}=1 / \sqrt{8\pi G}$ refers to the reduced Planck mass, and $\Lambda$ is the energy scale of PV term. The Latin indices are raised and lowered by the Kronecker delta symbol. Without loss of generality, we consider $\dot{\phi} < 0$.

We employ the Fourier and helicity decomposition of GWs in terms of the circular polarization tensors as, 
\begin{align}
h_{ij}(t,\mathbf{x}) 
= \sum_{A=\rm{R,L}} \int \frac{d^3\mathbf{k}}{(2\pi)^{3/2}} h^A_{\mathbf{k}}(t) e_{ij}^A(\mathbf{k}) e^{i \mathbf{k} \cdot \mathbf{x}} ~,
\end{align}
where R and L refer to the right- and left-handed polarizations, respectively. The circular polarization tensors $e_{ij}^A(\mathbf{k})$ are defined as $e^{R}_{ij} \equiv (e^{+}_{ij} + i e^{\times}_{ij} )/\sqrt{2}$ and $e^{L}_{ij} \equiv (e^{+}_{ij} - i e^{\times}_{ij} )/\sqrt{2}$, where $e^{+,\times}_{ij}$ are linear polarization tensors. Hence, $e_{ij}^A(\mathbf{k})$ satisfies $\epsilon^{abc} \hat{k}_a e_{cl}^A = i \lambda_A e^{A b}_{l}$, with $\lambda_{\rm{R},\rm{L}}= 1, -1$ showing the helicity dependence.
The mode function satisfies the following EoM from the action~\eqref{eq:action},
\begin{equation} \label{eq:eom_hk}
    \ddot{h}_{k}^{A}+3H\dot{h}_{k}^{A}+\frac{k}{a}\l( \frac{k}{a} + \lambda_A {\dot{\phi} \over \Lambda } \r) h_{k}^{A}=0 ~,
\end{equation}
where the term $\lambda_A\dot{\phi}/\Lambda$ manifests the parity violation.
From the last term in Eq.~\eqref{eq:eom_hk}, it is evident that the tachyonic instability on subhorizon scales, $(k/a)(k/a+\lambda_A\dot{\phi}/\Lambda)<0$ with $|\dot{\phi}/\Lambda| \gtrsim \mathcal{O}(H)$, can arise in $h_{k}^{R}$ (and $h_{k}^{L}$ for $\dot{\phi} > 0$). This occurs during reheating but does not happen during inflation, as constrained by the backreaction effect, as discussed below.
\footnote{
The PV term in~\eqref{eq:action} is similar to the chemical potential considered in Ref.~\cite{Tong:2022cdz}, which is used to enhance the cosmological collider signal. The chemical potential typically induces a frequency-dependent mass shift, making one particle mode easier to produce than the other. In our case, right-handed GWs are produced more readily than left-handed GWs.}

\begin{figure*}[ht]
    \centering
    \includegraphics[width=0.9\textwidth ]{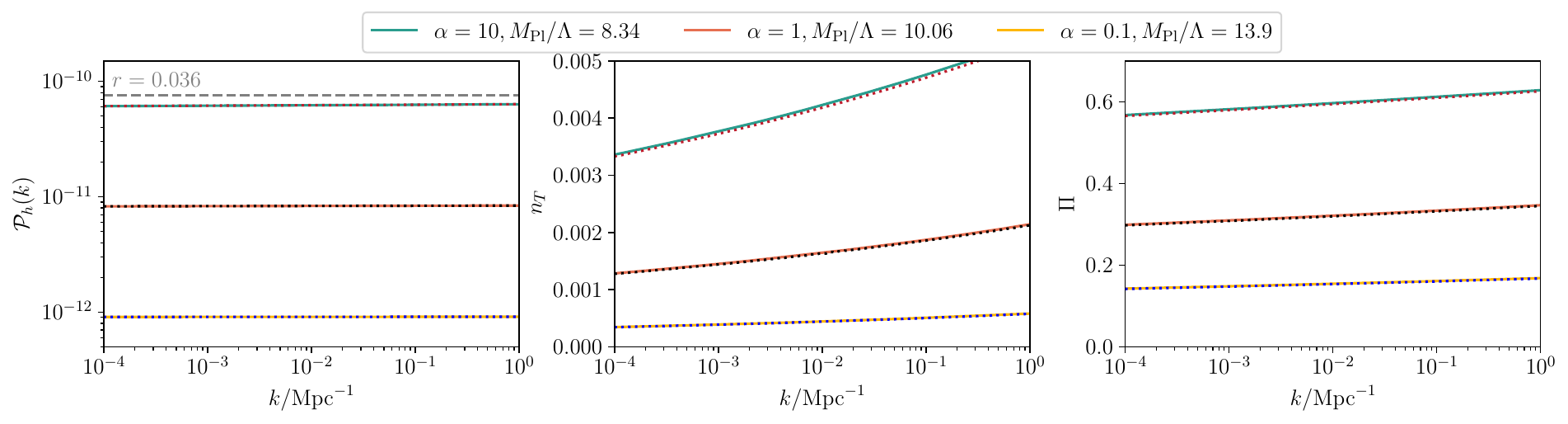}
    \caption{ The total power spectrum $\mathcal{P}_h$, spectral index $n_T$, and chirality $\Pi$ of primordial GWs at CMB scales for three cases: $(\alpha, M_{\rm Pl}/\Lambda) = (10, 8.34)$, $(1, 10.06)$, $(0.1, 13.9)$. The $e$-folding number $N_{\rm reh}$ is set to $7$ in all three cases. The solid curves represent the numerical results, while the dotted curves correspond the approximated expressions based on Eqs.~\eqref{eq:PA_inf} and \eqref{n_T}. The gray dashed line in the left panel shows the vanilla scale-invariant tensor spectrum with $r = 0.036$ for comparison.}
    \label{fig:Ph}
\end{figure*}

\begin{figure*}[ht]
\centering
\includegraphics[width=0.9\textwidth ]{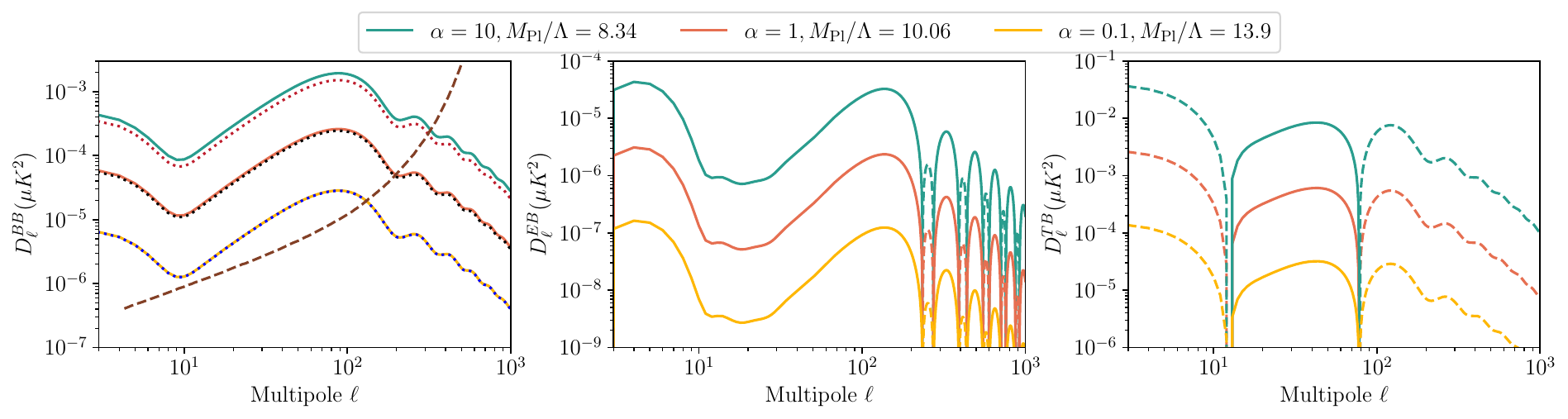}
\caption{The CMB angular power spectra, $D^{XY}_\ell = \ell (\ell +1 )C^{XY}_\ell/ (2\pi)$ with $XY=\{BB,EB,TB\}$, for three cases: $(\alpha, M_{\rm Pl}/\Lambda) = (10, 8.34)$, $(1, 10.06)$, $(0.1, 13.9)$. The $e$-folding number $N_{\rm reh}$ is set to $7$ in all three cases. 
The solid curves represent the numerical results for chiral GWs, while the dotted curves in the left panel represents BB spectra for standard GWs for comparison. The dashed brown curve represents the sensitivity curve of the LiteBIRD~\cite{Ishino:2016izb}.
The middle and right panels show the absolute values of spectra, and negative values denoted by dashed lines. 
}
\label{fig:eb_tb}
\end{figure*}


Without loss of generality, we consider the standard single-field slow-roll inflation, and the background evolution is given by
\begin{equation} \label{eq:eom_bg}
3M_{\rm Pl}^2H^2=\frac{1}{2}\dot\phi^2+V(\phi) ~, 
\quad
\ddot\phi + 3H\dot\phi+V_{,\phi}=0 ~,
\end{equation}
which satisfies the slow-roll conditions, $\epsilon\equiv-\dot H/H^2 =\dot\phi^2/(2M_{\rm Pl}^2H^2) \ll 1$ and $|\ddot\phi/(H\dot\phi)|\ll1$, during inflation. 
Imposing the Bunch-Davies vacuum as the initial condition and the de Sitter approximation $\tau \simeq -1/(aH)$ with the conformal time $\ddd\tau = \ddd t/a$, one can solve the EoM~\eqref{eq:eom_hk} analytically (see the Supplemental Material for details), 
\begin{align}\label{eq:sol_hk}
    h_{k}^{A}(\tau) = {\sqrt{2} \over \sqrt{k} M_{\rm Pl} a(\tau)} &e^{i\theta_1\l( \eta_A,x \r)} \l( -2ix \r)^{2+i\eta_A} \nonumber
    \\&\times U\l( 2+i\eta_A,4,-2ix \r) ~,
\end{align}
where $\eta_A\equiv\lambda_A \eta$ with $\eta=-\dot{\phi}/\l( 2H \Lambda \r)$, and $x\equiv-k\tau$. $U$ is known as the Kummer function of the second kind~\cite{olver2010nist}. The real function is defined as $\theta_1\l( \eta_A, x \r)= x - \eta_A \ln(2x)-\pi/2 + \rm{arg}~\Gamma(2+i\eta_A)$ with the Gamma function $\Gamma(z) \equiv \int_{0}^{\infty} e^{-t} t^{z-1} \ddd t$.
\footnote{Analogous outcomes were achieved within the Nieh-Yan modified teleparallel gravity theory~\cite{Li:2023fto}, employing the solutions of the Whittaker equation.}
The GW power spectrum for each helicity at horizon crossing $a_\ast H_\ast=k$ is calculated as
\begin{align} \label{eq:PA_inf}
\mathcal{P}_{h}^{A}\l( k \r) = \frac{k^3}{2\pi ^2}\l| h_{k}^{A} \r|_{\ast}^2
= A_{h\ast} e^{\pi \eta_{A\ast}} ~,
\end{align}
where $A_{h\ast} \equiv \frac{H_{\ast}^{2}}{\pi^2M_{\mathrm{Pl}}^{2}} \frac{\sinh( \pi \eta_{A\ast} )}{\pi \eta_{A\ast}( 1+\eta_{A\ast}^2 )} = \frac{H_{\ast}^{2}}{\pi^2M_{\mathrm{Pl}}^{2}} \frac{\sinh( \pi \eta_{\ast} )}{\pi \eta_{\ast}( 1+\eta_{\ast}^2 )}$, the subscript ``$\ast$'' indicates that a quantity is evaluated at horizon crossing. 
In the weak coupling limit, $|\eta| \ll 1$, the above power spectra reduce to the standard results predicted by slow-roll inflation, as expected. 
Equation~\eqref{eq:PA_inf} shows that $\mathcal{P}_{h}^{L} < H_{\ast}^{2}/(\pi^2M_{\mathrm{Pl}}^{2})<\mathcal{P} _{h}^{R}$. 
This difference is not due to tachyonic instability but rather arises from the distinct ``frozen time'' for each $k$ mode of $h_{k}^{R,L}$. From Eq.~\eqref{eq:eom_hk}, the PV term ${\dot{\phi}}/{(a \Lambda)} \epsilon^{ijk} h_{il}\partial_j h_{k}^l$ induces a smaller sound horizon for $h_{k}^{R}$ such that $h_{k}^{R}$ exit the horizon earlier than $h_{k}^{L}$ for each $k$ mode, the amplitude of $h_{k}^{R}$ is thus larger than $h_{k}^{L}$. 
The chirality of GWs can be measured by their ratio $\Pi$, defined as $\Pi \equiv (\mathcal{P}_{h}^{R}-\mathcal{P}_{h}^{L})/(\mathcal{P}_{h}^{R}+\mathcal{P}_{h}^{L})= \tanh \l( \pi \eta_\ast \r)$.
Since $\eta$ generally increases during the slow-roll inflation, both the total power spectrum $\mathcal{P}_h = \mathcal{P}_{h}^{R}+\mathcal{P} _{h}^{L}$ and the chirality $\Pi$ exhibit blue tilts at CMB scales. The tensor spectral index $n_T$ is calculated as
\begin{align}\label{n_T}
&n_T \equiv \frac{\ddd\ln{\mathcal{P}_h}}{\ddd\ln{k}}  =-2\epsilon_\ast + 
\nonumber \\
&\frac{-1-3\eta_\ast^2+\pi\eta_\ast(1+\eta_\ast^2)[\coth(\pi\eta_\ast)+\tanh(\pi\eta_\ast)]}{1+\eta_\ast^2} \frac{\dot\eta_\ast}{H_\ast\eta_\ast} ~.
\end{align}
Consequently, the standard consistency relation between $n_T$ and the tensor-to-scalar ratio $r$ is violated in our scenario.

Our numerical results by solving Eqs.~\eqref{eq:eom_hk} and \eqref{eq:eom_bg} are shown in Fig.~\ref{fig:Ph}, which clearly demonstrate the above qualitative discussions. 
We choose the E-model $\alpha$-attractor potential~\cite{Kallosh:2013yoa, Kallosh:2019jnl} as an example, $V(\phi) = V_0 [ 1 -\exp( -\sqrt{2/({3\alpha})} \phi ) ]^2$, which successfully realizes the vanilla single-field slow-roll inflation and an oscillation phase after inflation.

The predicted chiral GWs, in the aHz-fHz frequency band, not only induce BB spectrum, but also lead to the non-vanishing EB and TB spectra, which are calculated as~\cite{Saito:2007kt,Gluscevic:2010vv}, $C^{BB}_\ell = 4\pi \int \ddd(\ln k) \l[ \mathcal{P}^L_h(k) + \mathcal{P}^R_h(k)\r] \Delta^B_\ell(k) \Delta^B_\ell(k)$ and $C^{XB}_\ell = 4\pi \int \ddd(\ln k) \l[ \mathcal{P}^L_h(k) - \mathcal{P}^R_h(k)\r] \Delta^X_\ell(k) \Delta^B_\ell(k)$, where $X=\{T,E\}$ and $\Delta^{T/E/B}_\ell(k)$ are radiation transfer functions for $T/E/B$, respectively. Utilizing the publicly available Boltzmann code CLASS~\cite{Blas:2011rf}, we calculate BB, EB, and TB spectra, denoted as $D^{XY}_\ell = \ell (\ell +1 )C^{XY}_\ell/ (2\pi)$ with $XY=\{BB,EB,TB\}$, and present results in Fig.~\ref{fig:eb_tb}. It is clear that BB spectra are detectable in the future CMB experiment, LiteBIRD~\cite{Ishino:2016izb}. 
Moreover, if the tensor-to-scalar ratio $r$ is slightly below the current bound $0.036$, $\Pi$ will be detectable at the $1\sigma$ level if it exceeds $0.43$ for the CMBPol mission~\cite{CMBPolStudyTeam:2008rgp,Gluscevic:2010vv}, which corresponds to $\alpha = 10$ in our case. 

Since the strength of the tachyonic instability is inversely proportional to the energy scale $\Lambda$, there exists a lower limit for $\Lambda$ within a given inflationary background.
From Eq.~\eqref{eq:eom_hk}, the tachyonic instability grows exponentially with the inflaton's speed $|\dot{\phi}|$, the strong enhancement of GWs can happen in the early stage of reheating, imposing stringent constraint on $\Lambda$, as we will discuss below.


The inflaton begins oscillating around the minimum of its potential after inflation, marking the onset of reheating. We begin with a qualitative analysis as below.
The potential near its minimum is approximately quadratic, $V(\phi)\simeq m^2\phi^2/2$ wtih $m$ the effective mass of inflaton, the background dynamics is given by $a \sim t^{2/3}$ and $\phi \sim a^{-3/2}\cos mt$. If inflaton oscillations begin at an initial time $t_0$, then $a\simeq a_0 (t/t_0)^{2/3}$ with $a_0=1$. 
The corresponding solution for $\phi$-field can be written as $\phi \simeq \Phi a^{-3/2}\cos mt $ with $\phi(t_0)=\Phi$. After the redefinition $u_k^A=a^{3/2}h_k^A$, Eq.~\eqref{eq:eom_hk} can be approximately expressed as
\begin{equation} \label{eq:eom_uk}
    \ddot{u}_k^A + \l( \omega_k^{A}(t) \r)^2u_k^A = 0 ~,
\end{equation}
where
\begin{equation} \label{eq:omega_k}
    \l(\omega_k^{A}(t)\r)^2 \simeq \frac{k^2}{a^2}\l( 1 - \lambda_A \frac{m\Phi}{a^{1/2} \Lambda k} \sin mt \r) ~,
\end{equation}
using the approximation $H \ll m$ during reheating. 
It is clear that $\l(\omega_k^{A}(t)\r)^2$ oscillates over time with a decreasing amplitude, thereby triggering tachyonic instability when $\l(\omega_k^{A}(t)\r)^2 < 0$, followed by its eventual cessation as Universe expands. The alternating change in the sign of the inflaton’s velocity is expected to cause alternating amplification of each circular polarization. From Eq.~\eqref{eq:omega_k}, modes with very small or large $k$ only experience weak or no tachyonic instability, resulting in bump-like spectra seen in Figs.~\ref{fig:omega_reh} and~\ref{fig:omega_reh_N}.

\begin{figure}[ht]
    \centering
    \includegraphics[width= 1.0 \columnwidth  ]{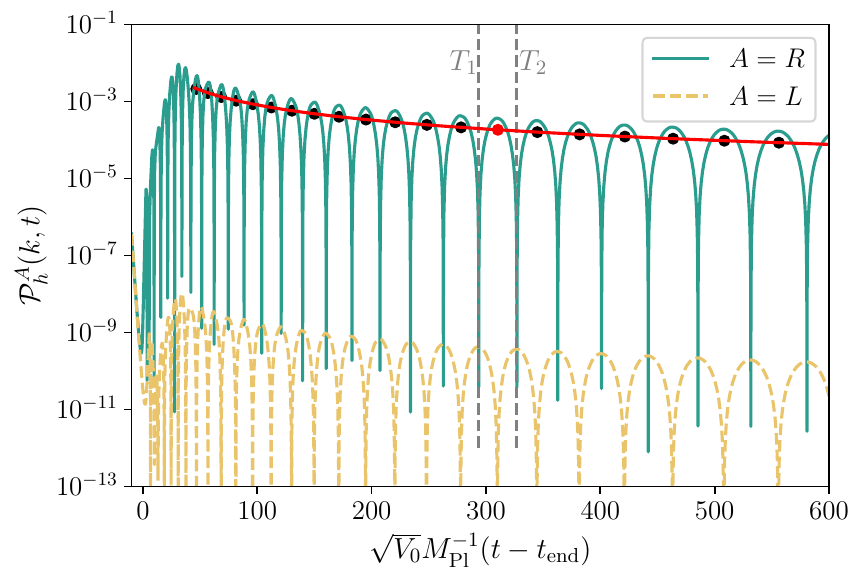}
    \caption{The chiral GW power spectra $\mathcal{P}^A_h(k,t)$ as a function of the dimensionless time variable $\sqrt{V_0}M_{\rm Pl}^{-1}(t-t_{\rm end})$ with $t_{\rm end}$ the end of inflation. The circle points represent the time-averaged values of $\mathcal{P}^R_h(k,t)$ over a single oscillation. The red curve depicts the approximation with parameters set by the red point.}
    \label{fig:ph_reh} 
\end{figure}

\begin{figure}[ht]
    \centering
    \includegraphics[width= 1.0 \columnwidth ]{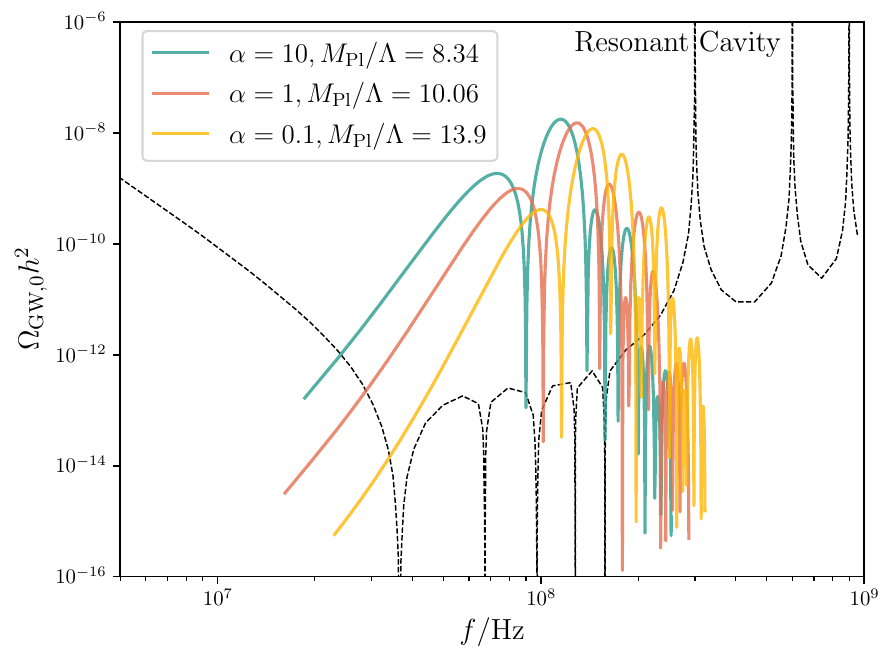}
    \caption{The current energy spectra of GWs produced during reheating for three cases: $(\alpha, M_{\rm Pl}/\Lambda) = (10, 8.34)$, $(1, 10.06)$, $(0.1, 13.9)$ that follow the condition~\eqref{eq:backreaction}. The $e$-folding number is chosen as $N_{\rm reh} = 7$ in all cases. The black dashed curve denotes the projected sensitivity of the resonant cavity experiment~\cite{Herman:2022fau}.}
    \label{fig:omega_reh}
\end{figure}

The full dynamics of chiral GWs during reheating can only be captured through numerical calculations. As a case study, we numerically solve Eqs.~\eqref{eq:eom_hk} and~\eqref{eq:eom_bg} during reheating, the chiral GW power spectra $\mathcal{P}_h^A(k,t)$ for a typical $k$ mode with $\alpha=1$ and $M_{\rm Pl}/\Lambda=10$ are shown in Fig.~\ref{fig:ph_reh}, which are consistent with the above qualitative analyses. 
After the growth regime, since $h_k^A \propto 1/a$ inside the horizon, we can express the time-averaged power spectrum, denoted by the overline, as $\overline{\mathcal{P}_h(k,t)} \simeq \overline{\mathcal{P}_h(k,t_\star)} \l[ {a(t)}/{a(t_\star)}\r]^{-2}$, where $\overline{\mathcal{P}_h(k,t_\star)} = \frac{1}{T_2-T_1}\int^{T_2}_{T_1} \mathcal{P}_h(k,t) \ddd t$, $t_\star=(T_2-T_1)/2$, and $T_1$, $T_2$ representing the start and end moments of a single oscillation of $\mathcal{P}_h(k,t)$, as illustrated by the red curve in Fig.~\ref{fig:ph_reh}.


The total energy density of chiral GWs is given by $\rho_{\rm GW}(t) = \frac{M_{\rm Pl}^2}{4} \sum_{A=L,R}\int\frac{\ddd k}{k}\frac{k^3}{2\pi^2}| \dot{h}^A_k |^2$.
We define $t_c$ as the moment when $\rho_{\rm GW}$ stops growing, ocurring after the initial oscillations of the inflaton, as also shown in Fig.~\ref{fig:ph_reh}. 
For simplicity, we require backreaction effects from the enhanced chiral GWs during reheating to be negligible,
\begin{equation} \label{eq:backreaction}
    {\rho_{\rm GW}(t_c) \over 3H^2M_{\rm Pl}^2} \lesssim 0.1 ~,
\end{equation}
which sets a lower bound on $\Lambda$.
The GW energy spectrum is calculated as~\cite{Maggiore:2007ulw}, $\Omega_{\rm GW}(k,t) = \frac{1}{3H^2M_{\rm Pl}^2} \frac{\ddd\rho_{\rm GW}}{\ddd\ln k}\simeq \frac{1}{12}\l( \frac{k}{a(t)H(t)} \r)^2 \overline{\mathcal{P}_h(k,t)}$. 
After reheating, the Universe transits to a radiation-dominated phase, where $\Omega_{\rm GW}(k,t)$ stops decaying. The current energy spectrum of GWs can be derived as 
\begin{equation}\label{omega_t0}
    \Omega_{\rm GW,0}(k) = 0.83\l( \frac{g_\ast}{10.75}\r)^{-1/3}\Omega_{r,0}\Omega_{\rm GW}(k,t_{\rm reh}) ~,
\end{equation}
where $\Omega_{r,0}$ is the current energy density parameter of radiation, with $t_{\rm reh}$ the time at the end of reheating, and $g_\ast = 106.75$ the effective number of relativistic degrees of freedom at $t_{\rm reh}$. The observed frequency relates to the comoving $k$ as $f=1.546\times10^{-15}(k/1~\rm{Mpc}^{-1})~{\rm Hz}$.
The observed GW energy spectrum is also influenced by reheating history, measured by the $e$-folding number $N_{\rm reh}$ and the equation of state parameter $w_{\rm reh}$ (see e.g., Refs.~\cite{Giovannini:1998bp, Figueroa:2019paj, Haque:2021dha, Chen:2024roo}).

\begin{figure}[ht]
    \centering
    \includegraphics[width= 1.0 \columnwidth ]{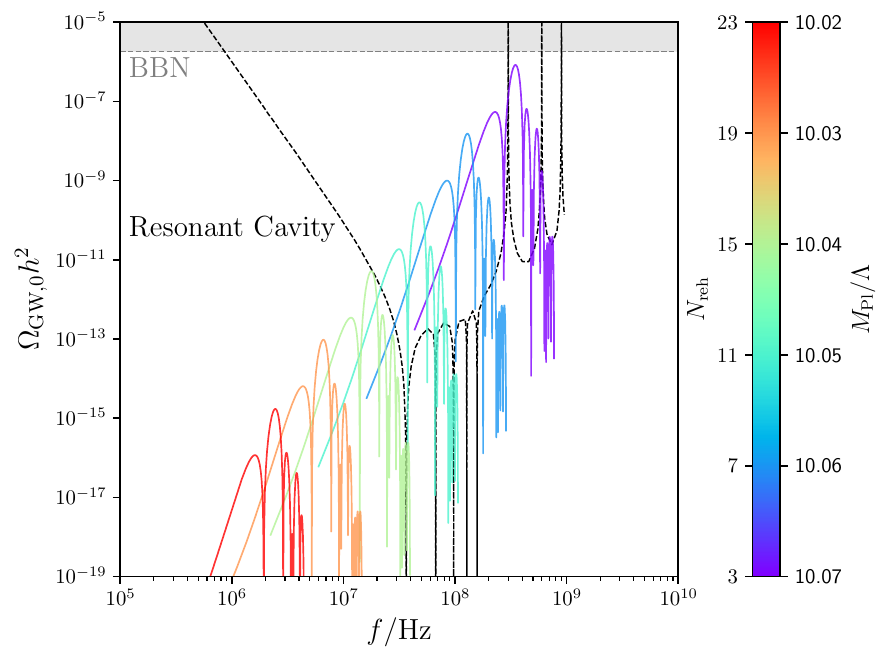}
    \caption{The current GW energy spectra generated during reheating with $\alpha=1$ for various $N_{\rm reh}$.  For each $N_{\rm reh}$, the energy scale $\Lambda$ is set to its lower limit that follows the condition~\eqref{eq:backreaction}. The gray-shaded region represents the bound from Big Bang Nucleosynthesis~\cite{Kohri:2018awv}. The black dashed curve denotes the projected sensitivity of a resonant cavity proposal~\cite{Herman:2022fau}. }
    \label{fig:omega_reh_N}
\end{figure}

As specific examples, we consider three cases, $\alpha = 10, 1, 0.1$, and then $M_{\rm Pl}/\Lambda = 8.34, 10.06, 13.9$ are chosen as the respective lower bounds that follow Eq.~\eqref{eq:backreaction}. We fix $N_{\rm reh} = 7$ in all cases\footnote{The $e$-folding number of inflation, $N(k)$, is related to $N_{\rm reh}$ as $N(k) = 67.2 -\ln {k \over 2.25 \times 10^{-4} {\rm Mpc}^{-1}} + {3 w_{\rm reh} - 1\over 4} N_{\rm reh} + \ln{H_k \over M_{\rm Pl}} - {1\over2} \ln{ H_{\rm end} \over  M_{\rm Pl}} $, where the Hubble parameter $H_k$ is evaluated at $k=aH$ and $H_{\rm end}$ denotes the Hubble parameter at the end of inflation~\cite{Liddle:2003as}. In this paper, we take the pivot scale as $k_{\rm p}=0.05~{\rm Mpc}^{-1}$.}, and present the observed GW energy spectra in Fig.~\ref{fig:omega_reh}, which may be probed by future resonant cavity experiments~\cite{Herman:2022fau}. 
The GW energy spectra for these cases differ in both peak frequency and shape, determined by inflaton's dynamics. The unique one-handed polarization and distinct energy spectrum profile of GW signals make them distinguishable from ultra-high frequency GW backgrounds originating from other sources, such as out-of-equilibrium excitation of scalar fields~\cite{Khlebnikov:1997di,Easther:2006vd,Garcia-Bellido:2007nns,Dufaux:2007pt,Figueroa:2017vfa,Fu:2017ero} and gauge fields \cite{Dufaux:2010cf,Figueroa:2016ojl,Adshead:2018doq,Adshead:2019igv,Bastero-Gil:2022fme} at preheating, and topological defects \cite{Gouttenoire:2019kij,Dror:2019syi,Buchmuller:2019gfy,Servant:2023tua}.
GWs generated during gauge-field preheating exhibit a distinct polarization pattern with amplification of either $-$ or $+$ polarization~\cite{Adshead:2018doq,Bastero-Gil:2022fme}, differing from circular polarizations in our scenario.
Moreover, the oscillating multiple-peak structure in the GW energy spectra originates from the dips in amplitudes at specific $k$ modes, which can be understood through the WKB approximation, see the Supplemental Material, which includes Refs.~\cite{Dufaux:2006ee,Abolhasani:2009nb,He:2020ivk,Zhu:2018smk}.

In order to show the effect of $N_{\rm reh}$ on $\Omega_{\rm GW,0}(k)$, we fix $\alpha=1$, and vary $N_{\rm reh}$ from $3$ to $23$, as shown in Fig.~\ref{fig:omega_reh_N}.
We observe that amplitudes of $\Omega_{\rm GW,0}(k)$ increase with decreasing $N_{\rm reh}$, since the potential during inflaton oscillations is approximately quadratic, the effective equation of state parameter is thus $w_{\rm reh} \simeq 0$, and the reheating energy density $\rho_{\rm reh}$ grows with respect to $\rho_{\rm GW}$. From Figs.~\ref{fig:omega_reh} and \ref{fig:omega_reh_N}, the backreaction condition~\eqref{eq:backreaction} imposes more stringent constraint than that from Big Bang Nucleosynthesis.

\section{Discussion}.
In this paper, we explore the generation of chiral GW signals driven by a PV term coupled to the inflaton, considering the entire evolution from inflation to reheating. Our results show that the resulting GWs are circularly polarized, spans a wide frequency range, and significantly enhanced at high frequencies. At CMB scales, chiral GWs contribute BB as well as non-vanishing EB and TB spectra in CMB, and can be tested by forthcoming CMB experiments, such as LiteBIRD. At high frequencies, particularly in the MHz-GHz range, GW signals exhibit a fully right-handed polarized state with a distinctive energy spectrum, which may be probed by future high-frequency GW detection concepts, including resonant cavity–based and quantum-enhanced proposals~\cite{Herman:2022fau,Kharzeev:2025lyu}. 
Our results provide a comprehensive understanding of chiral GWs across a broad frequency range, offering potential observational opportunities in both CMB and high-frequency regimes. These interrelated observations enable us to explore the PV interaction, the inflationary potential, and the reheating history.

Note that we do not focus on the energy spectrum of GWs at low frequencies $f < \rm{MHz}$. This is because inflation-generated GWs exhibit a nearly flat energy spectrum with a slight blue tilt from the inflaton's slowly increasing velocity. Even disregarding the constraint on $\Lambda$ from reheating, CMB observations impose an upper bound on the amplitude of GWs at CMB scales. Consequently, the energy spectrum of GWs remains suppressed at frequencies accessible to future GW experiments such as SKA~\cite{Janssen:2014dka}, LISA~\cite{Bartolo:2016ami}, and the Einstein Telescope~\cite{ET:2019dnz}, falling below their sensitivity curves.

This paper explores chiral GWs in the no-backreaction regime, however, investigating the backreaction of over-amplified GWs (corresponding to a lower energy scale $\Lambda$) during reheating is intriguing.
Although our conclusions on chiral GWs during inflation is independent of inflationary potentials, the $\alpha$-attractor potential that we adopted induces an oscillation phase with $w_{\rm reh} \simeq 0$, which affects the evolution of GWs after inflation. Hence, it is also worth examining how this equation-of-state parameter influences the predicted chiral GW signals. We leave these questions for future works.

\begin{acknowledgments} 

C.F is supported by the National Key Research and Development Program of China Grant No. 2020YFC2201502, and the National Natural Science Foundation of China Grant No. 12305057.
C.C and Y.W are supported in part by the National Key Research and Development Program of China (Grant No. 2021YFC2203100) and NSFC (Grants No. 12433002). C.F and C.C. thank Yu-Min Hu, Jia-Rui Li, Xi Tong, Dongdong Zhang, Tao Zhu and Wang-Wei Yu for stimulating discussions. C.C. thanks the support from the Jockey Club Institute for Advanced Study at The Hong Kong University of Science and Technology. 

\end{acknowledgments}

\appendix

\begin{widetext}

\subsection{Gravitational parity-odd terms for linear gravitational waves}

In Fig.~\ref{fig:pv_terms}, we show the various PV terms for the linear tensor perturbations that emerge from the PV gravity theories. The full quadratic tensor action containing these PV terms can be written as
\begin{align}
	S_{T}^{(2)}  = \frac{M_{\mathrm{Pl}}^2}{8} \int \ddd t \ddd^3x a^3(t) &\left[ \dot{h}_{ij}\dot{h}^{ij} -\frac{1}{a^2} \big( \partial h_{ij} \big)^2 
	+ {c_1(t) \over a M_{\mathrm{Pl}}} \epsilon^{ijk} \dot{h}_{i}^{l} \partial_j\dot{h}_{kl}
	+ {c_2(t) \over a^3 M_{\mathrm{Pl}}} \epsilon^{ijk} \partial^2h_{il} \partial_jh_{k}^{l} \right. \nonumber \\
	& \qquad\qquad\qquad\left. + {c_3(t) \over a} \epsilon^{ijk} \dot{h}_{il} \partial_jh_{k}^{l}
	+ {c_4(t) M_{\mathrm{Pl}} \over a} \epsilon^{ijk} h_{il}\partial_j h_{k}^l \right] ~,
\end{align}
which leads to the following EoM via the variation principle: 
\begin{align}
	\left[ 1 +  \lambda_A {c_1 \over M_{\mathrm{Pl}}} {k\over a} \right]  \ddot{h}_{k}^{A} + &\left[ 3 + \lambda_A \left( 2 {c_1 \over M_{\mathrm{Pl}}} + {\dot{c}_1 \over H M_{\mathrm{Pl}}} \right) {k \over a} \right] H\dot{h}_{k}^{A} 
	+ {k^2\over a^2} \left[ 1 +\lambda_A {c_2 \over M_{\mathrm{Pl}}} {k\over a} \right. \nonumber \\
	&  \left. \qquad\qquad\qquad + \lambda_A \left(c_3 H + {1\over2} \dot{c}_3 \right) \left({k \over a}\right)^{-1} 
	- \lambda_A c_4 M_{\mathrm{Pl}} \left({k \over a}\right)^{-1} \right] h_{k}^{A} = 0 ~.
\end{align}
Note that all coefficients $c_{1,2,3,4}(t)$ are dimensionless and are related to coefficients $f_{1,2,3,4}(t)$, shown in Fig.~\ref{fig:pv_terms}, through the following relations: 
\begin{align}
	f_1 = -{c_1 \over M_{\mathrm{Pl}}} ~,
	\quad 
	f_2 = - 2 {c_1 \over M_{\mathrm{Pl}}} - {\dot{c}_1 \over H M_{\mathrm{Pl}}}~, 
	\quad 
	f_3 = -{c_2 \over M_{\mathrm{Pl}}} ~, 
	\quad
	f_4 = - c_3 H - {1\over2} \dot{c}_3 ~, 
	\quad 
	f_5 = c_4 M_{\mathrm{Pl}} ~.
\end{align}

\begin{figure*}[ht]
	\centering
	\includegraphics[width = 0.45\textheight]{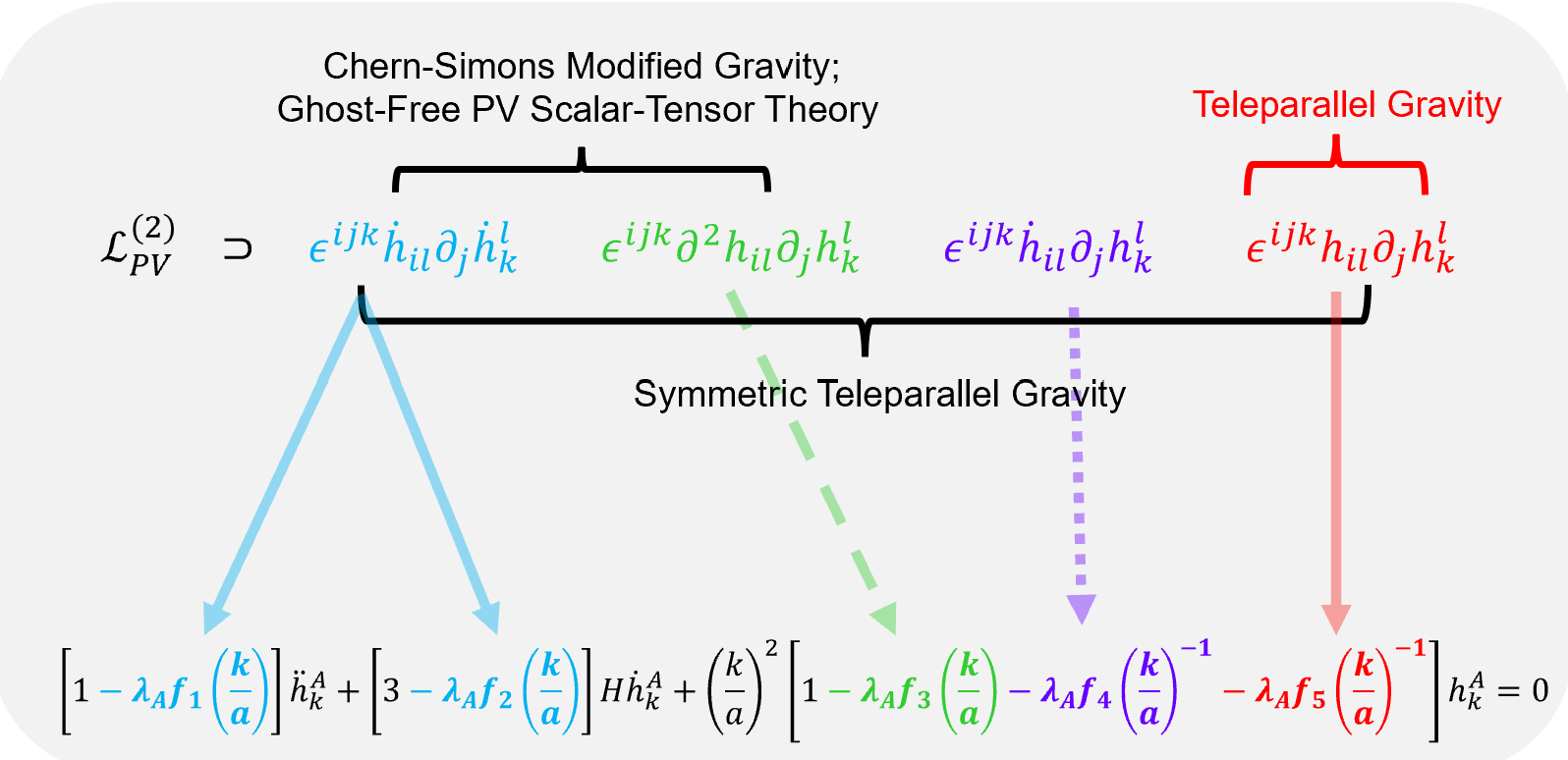}
	\caption{Various PV terms with different temporal and spatial derivatives for linear GWs in gravity theories, including Chern-Simons modified gravity~\cite{Jackiw:2003pm,Alexander:2009tp}, ghost-free PV scalar-tensor theory~\cite{Crisostomi:2017ugk,Nishizawa:2018srh,Gao:2019liu}, PV extensions of teleparallel gravity~\cite{Li:2020xjt,Li:2022mti} and symmetric teleparallel gravity~\cite{Conroy:2019ibo,Li:2021mdp,Li:2022vtn}.}
	\label{fig:pv_terms}
\end{figure*}

\subsection{Inflation}
\renewcommand{\theequation}{A.\arabic{equation}} 		
\setcounter{equation}{0}
\label{app:inf}

In order to solve the EoM~(3), we introduce the new variable $v_{k}^{A} \equiv {M_{\rm Pl}\over2} a h_{k}^{A}$, and it satisfies the following equation derived from Eq.~(3),
\begin{align} \label{eq:eom_vk}
	{\ddd^2 v_{k}^{A} \over \ddd x^2} + \l( 1 - \frac{2}{x^2} - \frac{2\eta _A}{x} \r) v_{k}^{A}=0 ~,
\end{align}
where we used the de Sitter approximation $\tau \simeq -1/(aH)$ in the derivation.
Given $\dot\phi < 0$ during inflation, $\eta$ can be expressed as $\eta = M_{\rm Pl} \sqrt{\epsilon/2}/\Lambda$. 
For simplicity, we treat $\eta$ as a constant, which is generally justifiable in the standard slow-roll inflation.
Therefore, the GW dynamics \eqref{eq:eom_vk} is identified as the Coulomb wave equation~\cite{olver2010nist}, of which the general solution is given by
\begin{align}\label{eq:solution}
	v_{k}^{A}(\tau) = C_1\l( k \r) H_{1}^{+}\l( \eta_A, x \r) +C_2\l( k \r) H_{1}^{-}\l( \eta_A, x \r),
\end{align}
where $H_{1}^{+}$ and $H_{1}^{-}$ are two independent solutions defined as
\begin{align}
	H_{1}^{\pm}\l( \eta_A, x \r) =e^{\pm i\theta_1\l( \eta_A, x \r)}\l( \mp 2 i x \r)^{2\pm i\eta_A} U\l( 2\pm i\eta_A,4,\mp 2ix \r) ~,
\end{align}
where $U(a, b, z)$ satisfies $z \ddd^2 U / \ddd z^2 + (b-z) \ddd U / \ddd z - a U = 0$, known as the Kummer function of the second kind. 
The coefficients $C_{1,2}(k)$ in \eqref{eq:solution} can be determined by the initial condition, setting deep inside the Hubble horizon, i.e., the Bunch-Davies vacuum solution $e^{ix}/\sqrt{2k}$ for $x \gg 1$. Considering the asymptotic form $H_{1}^{\pm} \rightarrow e^{\pm i x}$, we solve $C_1(k)=1/\sqrt{2k}$ and $C_2(k)=0$. Hence, we derived the time evolution of mode functions of chiral GWs during inflation as,
\begin{equation}
	v_{k}^{A}(\tau)=\frac{1}{\sqrt{2k}}e^{i\theta_1\l( \eta_A,x \r)}\l( -2ix \r)^{2+i\eta_A}U\l( 2+i\eta_A,4,-2ix \r) ~,
\end{equation}
from Eq. \eqref{eq:solution}.
The observable chiral GWs eventually exit the Hubble horizon during inflation, and considering $U\l(2+i\eta_A,4, -2ix\r) \rightarrow 2\l( -2ix \r)^{-3}/\Gamma\l( 2+i\eta_A \r)$ on superhorizon scale $x \ll 1$, we obtain,
\begin{align}
	v_{k}^{A}(\tau) \simeq \frac{1}{\sqrt{2k}}e^{i\theta_1\l( \eta_A,x \r)} \l( -2ix \r)^{-1+i\eta_A} \frac{2}{\Gamma \l( 2+i\eta_A \r)} ~.
\end{align}
Reverting to the mode function $h_k^A$, one can show that $|h_k^A(\tau)|$ remains constant on superhorizon scales similar to the standard GWs, 
\begin{equation} \label{eq:sol_hk}
	\l| h_{k}^{A} \r| \simeq {2\over M_{\rm Pl}} \frac{H}{\sqrt{2k^3}} \sqrt{\frac{\sinh \l( \pi \eta_A \r)}{\pi \eta_A\l( 1+\eta_A^2 \r)}} e^{\frac{\pi}{2} \eta_A} ~, 
\end{equation} 
where we used formulas $\Gamma \l( 2+i\eta_A \r) \Gamma( 2-i\eta_A ) = \pi \eta_A( 1+\eta_A^2 )/ \sinh( \pi \eta_A )$, $(-i)^i = e^{\pi/2}$, and $( -2ix )^{-1+i\eta_A} = {i\over2x} e^{i \eta_A \ln(2x)} e^{\pi \eta_A/2}$. Then, it is straightforward to derive the power spectrum~(6) using the above solution~\eqref{eq:sol_hk}.

\subsection{Reheating}
\renewcommand{\theequation}{B.\arabic{equation}} 		
\setcounter{equation}{0}
\label{app:reheating}

In what follows, we present analytical calculations of Eq.~(8) during reheating, building upon the WKB approximation outlined in Refs.~\cite{Dufaux:2006ee,Abolhasani:2009nb,He:2020ivk}\footnote{Another analytical method, called the uniform asymptotic approximation, of calculating particle production during reheating has been studied in Ref.~\cite{Zhu:2018smk}.}, {\it mutatis mutandis}. Our analyses provide valuable insights into characteristic features of GW energy spectra shown in Figs.~5 and~6.

\begin{figure}[ht]
	\centering
	\includegraphics[width=0.4\textwidth]{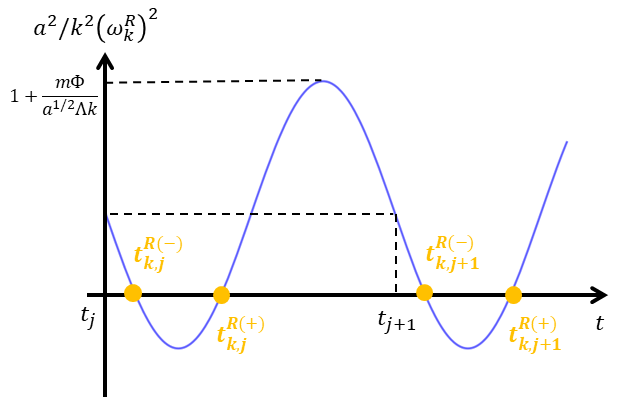}
	\includegraphics[width=0.4\textwidth]{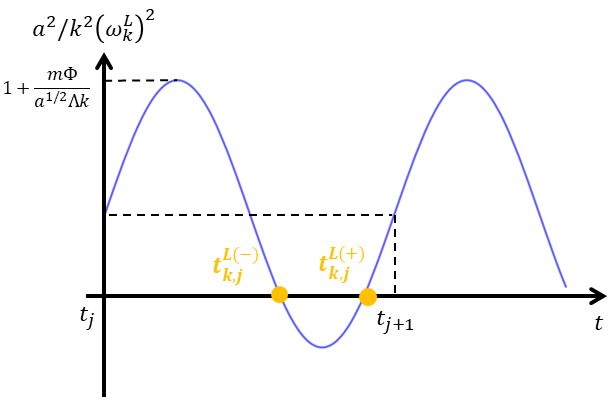}
	\caption{Sketches that depict squared frequencies, ${a^2\over k^2}(\omega_{k}^{A})^2$, for right- and left-handed of GWs during two successive oscillation periods. The turning points, $t_{k,j}^{A(-,+)}$, where $(\omega_{k}^{A})^2 =0$, are indicated by solid circles.}
	\label{fig:wkb}
\end{figure}

First, we need to examine the particle production due to one oscillation period. Considering the $(j+1)$-th period (namely $t_{j} \leq t \leq t_{j+1}$) as shown in Fig.~\ref{fig:wkb}, the WKB approximation of Eq.~(8) can be applied in three regions,
\begin{equation} \label{eq:wkb_uk}
	u^A_k(t) \simeq
	\l\{
	\begin{matrix}
		\frac{\alpha_{k,j}^A}{\sqrt{2\omega_k^{A}(t)}}\exp\l( -i\int_{t_{k,j}^{A(-)}}^{t} \omega_k^{A}(t')\ddd t'\r)
		+ \frac{\beta_{k,j}^A}{\sqrt{2\omega_k^{A}(t)}}\exp\l(i\int_{t_{k,j}^{A(-)}}^{t} \omega_k^{A}(t')\ddd t'\r) ~, & t_j \leq t < t_{k,j}^{A(-)} 
		\\ \\
		\frac{a_{k,j}^A}{\sqrt{2\Omega_k^{A}(t)}}\exp\l( -\int^t_{t^{A(-)}_{k,j}}\Omega_k^{A}(t')\ddd t'\r)
		+ \frac{b_{k,j}^A}{\sqrt{2\Omega_k^{A}(t)}}\exp\l(\int^t_{t^{A(-)}_{k,j}}\Omega_k^{A}(t')\ddd t'\r) ~, & t_{k,j}^{A(-)} < t < t_{k,j}^{A(+)}
		\\ \\
		\frac{C_{k,j}^A}{\sqrt{2\omega_k^{A}(t)}}\exp\l( -i \int_{t_{k,j}^{A(+)}}^{t} \omega_k^{A}(t')\ddd t'\r)
		+ \frac{D_{k,j}^A}{\sqrt{2\omega_k^{A}(t)}}\exp\l(i \int_{t_{k,j}^{A(+)}}^{t} \omega_k^{A}(t')\ddd t'\r) ~, & t_{k,j}^{A(+)} < t < t_{j+1} 
	\end{matrix}
	\r. 
	~,
\end{equation}
and we want to derive the transfer matrix between the Bogoliubov coefficients $(\alpha_{k,j}^A, \beta_{k,j}^A)$ and $(C_{k,j}^A, D_{k,j}^A)$ which satisfy $|\alpha_{k,j}^A|^2 - |\beta_{k,j}^A|^2 = |C_{k,j}^A|^2 - |D_{k,j}^A|^2 = 1$. Note that there exist two turning points $t_{k,j}^{A(-)}$, $t_{k,j}^{A(+)}$ ($\omega_k^{A} = 0$) where the above WKB approximation apparently breaks down, and we have to resort to another method to solve Eq.~(8). Linearizing $(\omega_k^{A})^2$ around turning points, 
\begin{equation} \label{eq:linearize}
	(\omega_k^{A}(t))^2 \simeq -g(t_{k,j}^{A}) (t - t_{k,j}^{A}) ~,
\end{equation}
where $g(t_{k,j}^{A}) \equiv -\ddd(\omega_k^{A}(t))^2/\ddd t \big|_{t_{k,j}^{A}}$ and $t_{k,j}^{A} = t_{k,j}^{A(-)}$ or $t_{k,j}^{A(+)}$. Then, Eq.~(8) can be reorganized in the form of an Airy equation,
\begin{equation}
	{\ddd^2 u(z)\over\ddd z^2} - z u(z) = 0 ~,
\end{equation}
where $z \equiv g^{1/3}(t-t_{k,j}^{A})$. Its solution is given by the linear combination of Airy functions,
\begin{equation} \label{eq:airy}
	u(z) = a {\rm Ai}(z) + b {\rm Bi}(z) ~,
\end{equation}
with unknown constants $a$ and $b$ which can be determined by the connection with WKB solutions in Eq.~\eqref{eq:wkb_uk} at turning points. In the large-$|z|$ limit, the asymptotic behaviors of Airy functions are given as~\cite{olver2010nist}
\begin{equation} \label{wkb:airy}
	{\rm Ai}(z) 
	\sim
	\l\{
	\begin{matrix}
		\pi^{-1/2} |z|^{-1/4} \sin(\xi+\pi/4) ~, & z \ll 0
		\\ \\
		{1\over2} \pi^{-1/2} |z|^{-1/4} e^{-\xi} ~, & z \gg 0
	\end{matrix}
	\r. ~,
	\quad
	{\rm Bi}(z)
	\sim
	\l\{
	\begin{matrix}
		\pi^{-1/2} |z|^{-1/4} \cos(\xi+\pi/4) ~, & z \ll 0
		\\ \\
		\pi^{-1/2} |z|^{-1/4} e^{\xi} ~,& z \gg 0
	\end{matrix}
	\r. ~,
\end{equation}
where $\xi \equiv {2\over3} |z|^{3/2}$.

Note that in the vicinity of $t_{k,j}^{A(-)}$, for $t<t_{k,j}^{A(-)}$ ($z < 0$), $\omega_k^{A}(t) = \sqrt{(\omega_k^{A}(t))^2} = g^{1/3} (-z)^{1/2}$, and $\int_{t}^{t_{k,j}^{A(-)}} \omega_k^{A}(t') \ddd t' = \xi$; For $t > t_{k,j}^{A(-)}$ ($z>0$), we have $\Omega_k^{A}(t) = \sqrt{(\Omega_k^{A}(t))^2} = g^{1/3} z^{1/2}$ and $\int_{t_{k,j}^{A(-)}}^{t} \Omega_k^{A}(t') \ddd t' = \xi$. Similarly, in the vicinity of $t_{k,j}^{A(+)}$, we have $\int_{t_{k,j}^{A(+)}}^{t} \Omega_k^{A}(t') \ddd t' = \xi$ for $t < t_{k,j}^{A(+)}$ ($z>0$) and $\int_{t}^{t_{k,j}^{A(+)}} \omega_k^{A}(t') \ddd t' = \xi$ for $t > t_{k,j}^{A(+)}$ ($z<0$). Using the WKB solutions~\eqref{eq:wkb_uk} and the asymptotic forms~\eqref{wkb:airy}, the matching conditions in the vicinity of $t_{k,j}^{A(-)}$ give
\begin{align}
	{a \over \pi^{1/2} |z|^{1/4}} \sin(\xi+\pi/4) + {b \over \pi^{1/2} |z|^{1/4}} \cos(\xi+\pi/4)
	&= 
	\frac{\alpha_{k,j}^A}{\sqrt{2\omega_k^{A}(t)}} e^{ i\xi }
	+ \frac{\beta_{k,j}^A}{\sqrt{2\omega_k^{A}(t)}} e^{ -i\xi } ~,
	\\
	{a \over 2\pi^{1/2} |z|^{1/4}} e^{-\xi} + {b \over \pi^{1/2} |z|^{1/4}} e^{\xi}
	&= {a_{k,j}^A \over \sqrt{2\Omega_k^{A}(t)}} e^{-\xi}
	+ {b_{k,j}^A \over \sqrt{2\Omega_k^{A}(t)}} e^{\xi} ~,
\end{align}
which lead to the following relationship,
\begin{equation} \label{wkb:trsfer_1}
	\begin{pmatrix}
		a_{k,j}^A \\ \\ b_{k,j}^A
	\end{pmatrix}
	=
	\begin{pmatrix}
		{1\over2} e^{i {\pi \over 4}} & {1\over2} e^{-i {\pi \over 4}}
		\\ \\
		e^{-i {\pi \over 4}} & e^{i {\pi \over 4}}
	\end{pmatrix}
	\begin{pmatrix}
		\alpha_{k,j}^A \\ \\ \beta_{k,j}^A
	\end{pmatrix} ~.
\end{equation}
Similarly, the matching conditions in the vicinity of $t_{k,j}^{A(+)}$ lead to the transfer matrix,
\begin{equation} \label{wkb:trsfer_2}
	\begin{pmatrix}
		a_{k,j}^A \\ \\ b_{k,j}^A
	\end{pmatrix}
	=
	\begin{pmatrix}
		e^{\gamma} e^{i {\pi \over 4}} & e^{\gamma} e^{-i {\pi \over 4}}
		\\ \\
		{1\over2} e^{-\gamma} e^{-i {\pi \over 4}} & {1\over2} e^{-\gamma} e^{i {\pi \over 4}}
	\end{pmatrix}
	\begin{pmatrix}
		C_{k,j}^A \\ \\ D_{k,j}^A
	\end{pmatrix} ~,
\end{equation}
where
\begin{equation}
	\gamma^A_{k,j} \equiv \int_{t_{k,j}^{A(-)}}^{t_{k,j}^{A(+)}} \Omega_k^{A}(t')\ddd t' > 0
\end{equation}
is a measure of the ``width'' of the tachyonic instability band (i.e., $(\omega_k^{A})^2 < 0$) during a single oscillation period. 
Combining two transformations~(\ref{wkb:trsfer_1}, \ref{wkb:trsfer_2}), the final transfer matrix is calculated as
\begin{equation} \label{wkb:trsfer_3}
	\begin{aligned}
		\begin{pmatrix}
			C_{k,j}^A \\ \\ D_{k,j}^A
		\end{pmatrix}
		= 
		\begin{pmatrix}
			{e^{-\gamma^A_{k,j}}\over4} + e^{\gamma^A_{k,j}} & - i \l( {e^{-\gamma^A_{k,j}}\over4} - e^{\gamma^A_{k,j}} \r)
			\\ \\
			i \l( {e^{-\gamma^A_{k,j}}\over4} - e^{\gamma^A_{k,j}} \r) & {e^{-\gamma^A_{k,j}}\over4} + e^{\gamma^A_{k,j}}
		\end{pmatrix}
		\begin{pmatrix}
			\alpha_{k,j}^A \\ \\ \beta_{k,j}^A
		\end{pmatrix}~,
	\end{aligned}
\end{equation}
which implies the Bogoliubov transformation, 
\begin{equation}
	|C_{k,j}^A|^2 - |D_{k,j}^A|^2 = |\alpha_{k,j}^A|^2 - |\beta_{k,j}^A|^2 = 1 ~,
\end{equation}
as expected from the time-independent Wronskian of the mode equation~(8).

Secondly, we need to match $(C_{k,j}^A, D_{k,j}^A)$ with the followed oscillation period, namely with the coefficients $(\alpha_{k,j+1}^A, \beta_{k,j+1}^A)$ which is defined similar to the solution~\eqref{eq:wkb_uk}, 
\begin{equation} \label{eq:wkb_uk_next}
	u^A_k(t) \simeq
	\frac{\alpha_{k,j+1}^A}{\sqrt{2\omega_k^{A}(t)}}\exp\l( -i\int_{t_{k,j+1}^{A(-)}}^{t} \omega_k^{A}(t')\ddd t'\r)
	+ \frac{\beta_{k,j+1}^A}{\sqrt{2\omega_k^{A}(t)}}\exp\l(i\int_{t_{k,j+1}^{A(-)}}^{t} \omega_k^{A}(t')\ddd t'\r) ~, 
	\quad
	t_{j+1} \leq t < t_{k,j+1}^{A(-)} ~.
\end{equation}
Observing that the solution for $t_{k,j}^{A(+)} < t < t_{j+1}$, shown in Eq.~\eqref{eq:wkb_uk}, is also applicable in the range $t_{j+1} \leq t < t_{k,j+1}^{A(-)}$. Combining the solutions~(\ref{eq:wkb_uk}, \ref{eq:wkb_uk_next}), we derive the transfer matrix as,
\begin{equation} \label{wkb:trsfer_4}
	\begin{pmatrix}
		\alpha_{k,j+1}^A \\ \\ \beta_{k,j+1}^A
	\end{pmatrix}
	=
	\begin{pmatrix}
		e^{-i \sigma^A_{k,j}} & 0
		\\ \\
		0 & e^{i \sigma^A_{k,j}}
	\end{pmatrix}
	\begin{pmatrix}
		C_{k,j}^A \\ \\ D_{k,j}^A
	\end{pmatrix} ~,
\end{equation}
where
\begin{equation}
	\sigma^A_{k,j} \equiv \int_{t_{k,j}^{A(+)}}^{t_{k,j+1}^{A(-)}} \omega_k^{A}(t')\ddd t' > 0
\end{equation}
measures the ``width'' of the stable band (i.e., $(\omega_k^{A})^2 > 0$).
Using the transformations~(\ref{wkb:trsfer_3}, \ref{wkb:trsfer_4}), we relate $(\alpha_{k,j}^A, \beta_{k,j}^A)$ with $(\alpha_{k,j+1}^A, \beta_{k,j+1}^A)$ as
\begin{equation} \label{eq:wkb_matrix}
	\begin{pmatrix}
		\alpha_{k,j+1}^A \\ \\ \beta_{k,j+1}^A
	\end{pmatrix}
	=
	\begin{pmatrix}
		\l( {e^{-\gamma^A_{k,j}}\over4} + e^{\gamma^A_{k,j}} \r) e^{- i \sigma^A_{k,j}} & - i \l( {e^{-\gamma^A_{k,j}}\over4} - e^{\gamma^A_{k,j}} \r) e^{- i \sigma^A_{k,j}}
		\\ \\
		i \l( {e^{-\gamma^A_{k,j}}\over4} - e^{\gamma^A_{k,j}} \r) e^{i \sigma^A_{k,j}}  & \l( {e^{-\gamma^A_{k,j}}\over4} + e^{\gamma^A_{k,j}} \r) e^{ i \sigma^A_{k,j}} 
	\end{pmatrix}
	\begin{pmatrix}
		\alpha_{k,j}^A \\ \\ \beta_{k,j}^A
	\end{pmatrix} ~.
\end{equation}
It is straightforward to check the Bogoliubov transformation,
\begin{equation}
	|\alpha_{k,j+1}^A|^2 - |\beta_{k,j+1}^A|^2
	= |\alpha_{k,j}^A|^2 - |\beta_{k,j}^A|^2
	= 1 ~.
\end{equation}
By applying the formula~\eqref{eq:wkb_matrix} iteratively, one can derive the particle number density $|\beta_{k,j+1}^A|^2$ produced after $j$ oscillations, based on the initial particle number density $|\beta_{k,1}^A|^2$.

Notably, for the strong tachyonic instability $\gamma^A_{k,j} \gg 1$, we have
\begin{equation}
	|\beta_{k,j+1}^A|^2 \simeq e^{2 \gamma^A_{k,j} }\left|\beta^A_{k,j}-i\alpha^A_{k,j}\right|^2 \simeq 4 e^{2 (\gamma^A_{k,j} +\gamma^A_{k,j-1}) } \left|\beta^A_{k,j-1}-i\alpha^A_{k,j-1}\right|^2\left(\cos\sigma^A_{k,j-1}\right)^2 ~.
\end{equation}
Hence, certain modes with negative frequency squared do not exhibit strong tachyonic growth (see an example shown in Fig.~\ref{fig:ph_reh_0}) due to the condition $\cos\sigma^A_{k,j-1} \simeq 0$, which accounts for the dips in GW energy spectra shown in Figs.~5 and~6. 

\begin{figure*}[ht]
	\centering
	\includegraphics[width=0.85\textwidth ]{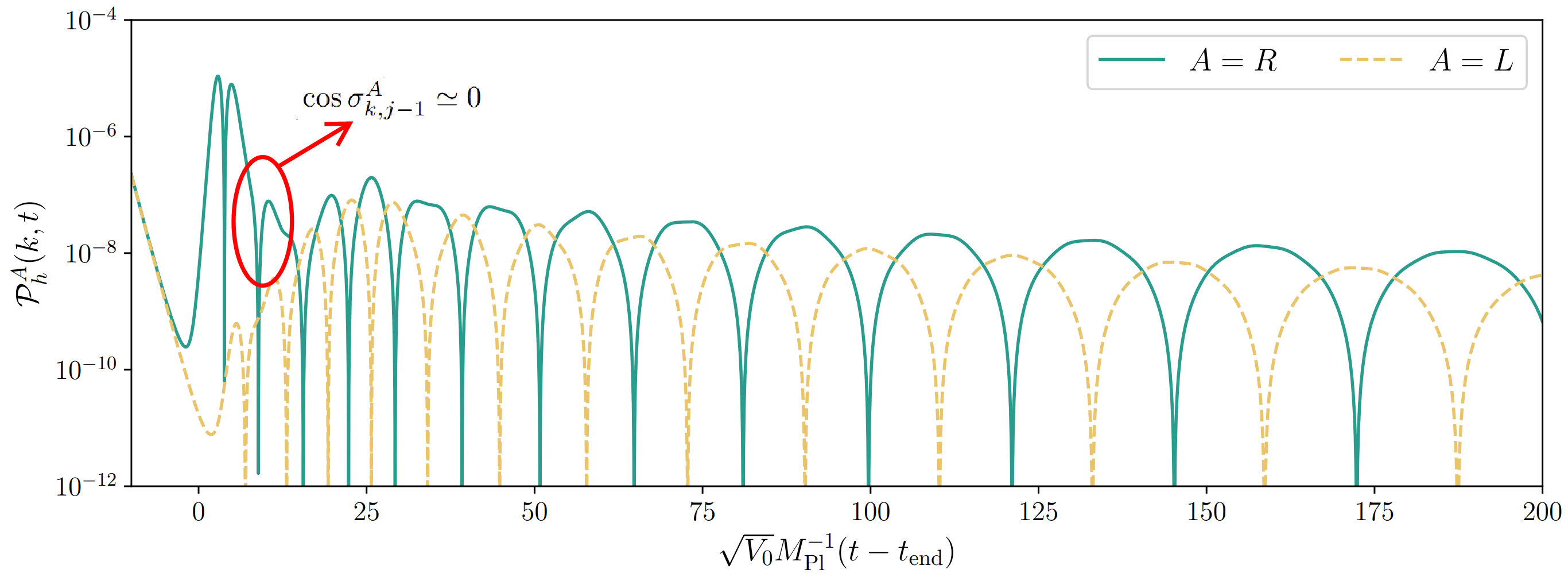}
	\caption{The evolution of $\mathcal{P}^A_h(k,t)$ for a typical $k$ mode do not exhibit strong tachyonic growth when $\cos\sigma^A_{k,j-1} \simeq 0$, as indicated by the red circle.}
	\label{fig:ph_reh_0} 
\end{figure*}

\end{widetext}

\bibliographystyle{apsrev4-1}
\bibliography{references}

\end{document}